\documentclass[5p,twocolumn]{elsarticle}

\usepackage{lineno,hyperref}
\modulolinenumbers[5]

\usepackage[english]{babel}
\usepackage{siunitx}
\usepackage{graphicx}
\usepackage{color}

    \makeatletter
    \def\ps@pprintTitle{%
       \let\@oddhead\@empty
       \let\@evenhead\@empty
       \def\@oddfoot{\reset@font\hfil\thepage\hfil}
       \let\@evenfoot\@oddfoot
    }
    \makeatother

\begin{document}

\begin{frontmatter}

\title{Investigating the Spectral Anomaly with Different Reactor Antineutrino Experiments}

\author[]{C.~Buck}

\author[]{A.P.~Collin}

\author[]{J.\ Haser\corref{corauth}}
\ead{julia.haser@mpi-hd.mpg.de}

\author[]{M.~Lindner}

\address{Max-Planck-Institut f\"{u}r Kernphysik, Saupfercheckweg 1, D-69117 Heidelberg, Germany}
\cortext[corauth]{Corresponding author}

\begin{abstract}
The spectral shape of reactor antineutrinos measured in recent experiments shows anomalies in comparison to neutrino reference spectra. New precision measurements of the reactor neutrino spectra as well as more complete input in nuclear data bases are needed to resolve the observed discrepancies between models and experimental results. This article proposes the combination of experiments at reactors which are highly enriched in ${}^{235}$U with commercial reactors with typically lower enrichment to gain new insights into the origin of the anomalous neutrino spectrum. The presented method clarifies, if the spectral anomaly is either solely or not at all related to the predicted ${}^{235}$U spectrum. Considering the current improvements of the energy scale uncertainty of present-day experiments, a significance of three sigma and above can be reached.
As an example, we discuss the option of a direct comparison of the measured shape in the currently running Double Chooz near detector and the upcoming Stereo experiment. A quantitative feasibility study emphasizes that a precise understanding of the energy scale systematics is a crucial prerequisite in recent and next generation experiments investigating the spectral anomaly.
\end{abstract}


\end{frontmatter}

\begin{figure}
 \centering
 \includegraphics[width=0.5\textwidth]{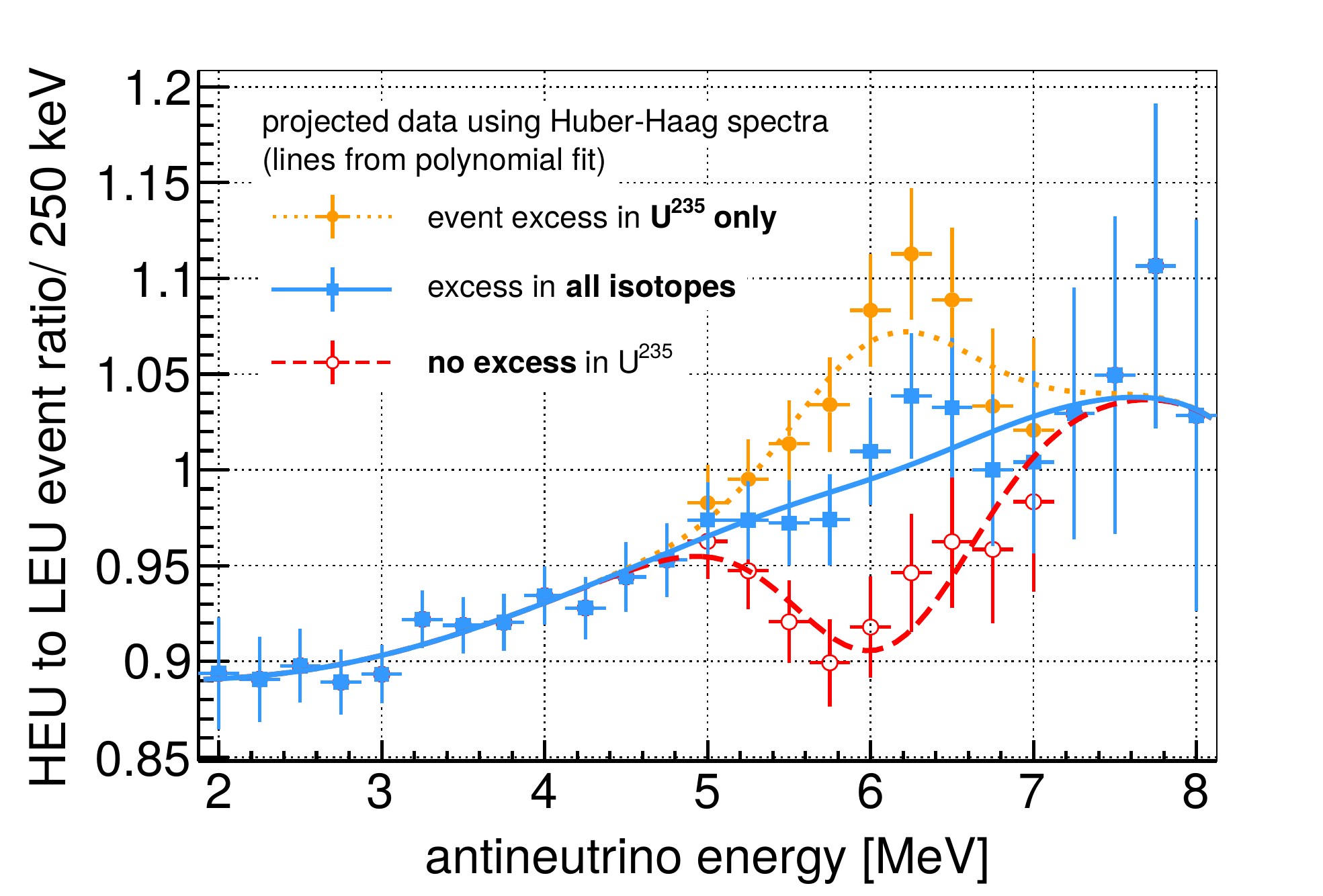}
 \caption{Event ratio of HEU to LEU antineutrino spectra for three hypotheses (see text). Data points show the event ratio of the projected data for Stereo (HEU) and Double Chooz near detector (LEU) using the Huber-Haag spectra for two years of data taking. The errors are statistical and include the model uncertainty of the Huber-Haag spectra taking into account correlations between the different isotopes. The detector response is not included in this plot.
 The lines are obtained from a polynomial fit to the Huber-Haag spectra.}
 \label{fig1}
\end{figure}

Intense research in the past few years has brought new insights in the antineutrino spectra emitted by nuclear reactors. Direct measurements of the antineutrino spectra as well as their improved predictions were a product of the search for the non-zero neutrino mixing angle $\theta_{13}$ at the km-baseline reactor experiments Double Chooz~\cite{PhysRevLett.108.131801}, Daya Bay~\cite{PhysRevLett.108.171803} and RENO~\cite{PhysRevLett.108.191802}. Although the experiments were successful at determining the $\theta_{13}$ parameter, the comparison of the measured spectra of km and short-baseline experiments to the most up-to-date predictions showed discrepancies both in absolute flux and spectral shape.
The inconsistency in total flux has a statistical significance of $2.7\,\sigma$ and is known as the reactor antineutrino anomaly~\cite{PhysRevD.83.073006}.
The energy shape distortion mainly manifests itself as a shoulder in the detected spectra at $E_{\nu}\!\sim\!\SI{6}{MeV}$ antineutrino energy~\cite{DCIII}.
\\
To date it is unknown if these differences indicate unaccounted physics in the propagation and detection of neutrinos. On the other hand, they could be introduced by the computational methods, theoretical assumptions or incomplete data inputs used to yield the predicted spectra. Furthermore, the two anomalies are in general treated as separate observations possibly caused by independent effects.
\\
Upcoming experiments will test if the discovered overall deficit in antineutrino rate is linked to neutrino flavour oscillations into a light sterile state~\cite{Abazajian:2012ys}. Being a possible explanation for the flux puzzle, this quantum mechanical phenomenon, however, does not explain the shoulder in the antineutrino spectra. This article will discuss the potential of current and future reactor neutrino experiments to resolve if the spectral shape distortion around 6~MeV is solely created by an incorrect prediction of the ${}^{235}$U spectral shape or if other actinides contribute with similar strength or even stronger. Knowledge could be gained by combining spectra from different reactor types.
\\\\
Nuclear reactors represent intense and extremely pure sources of electron antineutrinos with energies extending up to about \SI{10}{MeV}. Commercial power reactors are fueled with Low Enriched Uranium (LEU). Only a few percent of the fissile ${}^{235}$U is contained in the initial reactor fuel. During operation, more than 99\,\% of the emitted antineutrino flux is created by $\beta$-decays of the fission products of ${}^{235}$U, ${}^{238}$U, ${}^{239}$Pu and ${}^{241}$Pu. At present, the commonly used antineutrino reference spectra are from the Huber conversion~\cite{PhysRevC.84.024617} plus ${}^{238}$U from either the Mueller et al.~\cite{PhysRevC.83.054615} computation or the conversion of the Haag et al.~\cite{PhysRevLett.112.122501} measurement. The spectra are then often referred to as based on Huber-Mueller or Huber-Haag, respectively.
\\\\
Improvements on the performance of reactor antineutrino experiments at km-baselines were reached in the recent past, including an enhancement in energy resolution, energy scale uncertainty and an accumulation of higher statistics data.
These achievements allowed to accomplish precision measurements of the reactor antineutrino spectra and surprisingly a shoulder was found at $E_{\nu}\!\sim\!\SI{6}{MeV}$. When the ratio of measured and predicted spectrum is built, this shoulder becomes a ``bump'' in this energy region. The excess of events measured in the shoulder region of 5-7$\,\si{MeV}$ was found to be significant at $3\,\sigma$ and correlated to the thermal power of the reactors~\cite{DCIII}. Currently it is considered to be a common feature in all high-precision measurements of the antineutrino energy spectrum at nuclear reactors~\cite{Seo:2016uom,PhysRevLett.116.061801}, however with different magnitude of the effect.
As a consequence, there is widespread expectation that the spectral distortion is linked to inaccurate antineutrino reference spectra. This assumption is supported by thorough studies, as in Ref.~\cite{DCIII}, which question explanations of the shoulder being due to unaccounted backgrounds or detector response. However, owing to limited calibration data in the energy region of interest, a common bias in the non-linearity modelling, e.g. from approximations and simplifications in the quenching model of Birks~\cite{birks,1748-0221-6-11-P11006}, should not be fully excluded yet. Double Chooz, Daya Bay and RENO rely not only on a similar detector design but also the same inverse beta decay (IBD) detection technique, all using Gd doped liquid scintillators. Therefore independent confirmation of the spectral distortion with a different detector technology would be desirable.
\\\\
Besides using the reference spectra discussed earlier to build the flux and spectral shape prediction of nuclear reactors, a database depended computation of the prediction can be performed. This method, known as summation spectra, allows for detailed studies of the spectrum and its contributions from each fission product. Studies of the summation spectra pointed out that only few isotopes appear to contribute to the antineutrino spectrum above $\sim\,\SI{5}{MeV}$, predominantly ${}^{96}$Y and ${}^{92}$Rb~\cite{PhysRevLett.114.012502,Zakari-Issoufou:2015vvp}.
However, the uncertainties on summation spectra, are known to be sizeable~\cite{PhysRevD.83.073006}, if not even by far underestimated as suggested by Ref.~\cite{PhysRevD.92.033015}, where the authors use different fission yield databases to evaluate reactor spectra based on the summation method. Recent studies have shown that biases in the fission yield databases might cause the spectral structure, which implies that the databases need to be revised~\cite{Sonzogni2015AAP}.
\\
In addition, Hayes et al.~\cite{PhysRevD.92.033015} find hints for ${}^{238}$U being one of the actinides that might contribute significantly to the shoulder. The authors identify a possible correlation of the ${}^{238}$U content in the fuel of Double Chooz, Daya Bay and RENO and the relative overshoot of the spectrum in the bump region.
\\
In general summation spectrum calculations suffer from a lack of experimental data, which is replaced by theoretical assumptions, introducing sizeable uncertainties. 
Therefore experimental data from different reactor types could bring valuable insights into the nature of the reactor shape distortion bypassing the use of summation spectra and accordingly their large uncertainties.
\\\\
There are three large-scale experiments currently measuring the antineutrino flux emitted at nuclear reactors and which are suitable to explicitly study the shape of the detected energy spectrum: Double Chooz~\cite{DCIII}, Daya Bay~\cite{An:2015nua} and RENO~\cite{PhysRevLett.108.191802}. Each of the experiments relies on the IBD reaction ($\bar{\nu}_e + p \rightarrow e^{+} + n$) as detection mechanism, in which electron antineutrinos interact with free protons in form of hydrogen nuclei. Calorimetric measurement of the energy $E_\mathrm{visible}$ deposited by the created positron allows to derive the kinetic energy of the incident antineutrino via $E_\nu \approx E_\mathrm{visible} + \SI{0.8}{MeV}$. In Table~\ref{table1} the key parameters of these three reactor neutrino experiments are given.
The new generation of antineutrino detectors placed at nuclear reactor sites are summarized in Table~\ref{table2}. These experiments will search for light sterile neutrinos imprinting an unambiguous oscillation pattern with a $L/E_\nu \approx \SI{1}{m/MeV}$ oscillation signature in the measured event rate. The required experimental conditions are naturally given at research reactors, which are in most cases loaded with HEU (Highly Enriched Uranium) fuel which simplifies the computation of the predicted flux and spectrum owing to the absence of burn-up effects. Hence, for a core operated with HEU, it is a good approximation that the produced antineutrinos are exclusively generated by $\beta$-instable ${}^{235}$U fission daughters.
The small size of the sterile mixing angle~\cite{KOPP2013} requires detectors to possess systematics on the percent level as well as high statistics of the signal events. When it comes to the expected daily IBD rate the lower target mass or moderate reactor power -- compared to the experiments of Table~\ref{table1} -- is compensated by the shorter detector to reactor distance. Regarding the energy resolution, however, most of these experiments are inferior to the km-baseline detectors. Because of shallow depths of the reactor buildings and thus the experimental sites, the cosmic background flux is barely reduced. Nevertheless, for most of these experiments the main source of background originates from the reactors themselves, in form of gamma radiation and neutrons~\cite{PhysRevD.93.112006}.
\\\\
For the analysis proposed in this article the Stereo project is used as example. Benefitting from the good energy resolution and signal-to-background ratio at the same time, it is currently under construction and data taking will start end of 2016. Located at a research reactor highly enriched in ${}^{235}$U at ILL Grenoble, France, it allows to measure the antineutrino spectrum generated by the fission products of ${}^{235}$U.
\begin{table}
  \centering
  \caption{Near detector parameters of currently running $\theta_{13}$-experiments~\cite{DCIII,An:2015nua,PhysRevLett.108.191802}. The target mass $m_\mathrm{t}$, thermal power of the nuclear reactor $P_\mathrm{th}$ and the flux-weighted baseline $L$ are given. $R_\mathrm{IBD}$ denotes the average IBD rate. The energy resolution $\sigma_E/E$ is given at \SI{1}{MeV} visible energy. For the case of Double Chooz the resolution was inferred from the far detector performance.}
  \label{table1}
  \begin{tabular}{l >{\centering}p{2cm}<{\centering} p{1.7cm}<{\centering} p{1.5cm}<{\centering}}
  \hline\hline
      &  Double Chooz  &  Daya Bay   &  RENO  \\
  \hline
  $m_\mathrm{t}$ [t]  				& 8.3 & $4\times20$ & 15 \\
  $P_\mathrm{th}$ [$\mathrm{GW}$]  			& $2\times4.25$ & $6\times2.9$ & $6\times2.8$ \\
  $L$ [m]  					& 400 & 500 & 409 \\
  $R_\mathrm{IBD}$ [$\mathrm{day}^{-1}$]		& 300 & 1900 & 780 \\
  $\sigma_E/E$ 					& 0.08 & 0.08 & 0.07 \\
  \hline\hline
  \end{tabular}
\end{table}
\begin{table*}
  \centering
  \caption{Upcoming short baseline projects at nuclear reactors and their projected parameters. The IBD detection technique (PS: plastic scintillator, LS: liquid scintillator), the target mass $m_\mathrm{t}$, thermal power of the nuclear reactor $P_\mathrm{th}$ and the reactor to detector baseline $L$ are given. $R_\mathrm{IBD}$ denotes the expected IBD rate at reactor on and shortest baseline, S/B is the signal-to-background ratio. The photon statistical part of the energy resolution $\sigma_{E,\mathrm{Ph}}/E$ is given at \SI{1}{MeV} visible energy. The reactor fuel is classified in Low Enriched Uranium (LEU) or Highly Enriched Uranium (HEU).}
  \label{table2}
  \begin{tabular}{l >{\centering}p{2cm}<{\centering} >{\centering}p{1.1cm}<{\centering} p{1.5cm}<{\centering} p{1.1cm}<{\centering} p{1.7cm}<{\centering} p{1.5cm}<{\centering} p{1.1cm}<{\centering} p{1.7cm}<{\centering}}
  \hline\hline
  experiment [Ref.]	& technology &  $m_\mathrm{t}$ [t]  &  $P_\mathrm{th}$ [$\mathrm{MW}$]  &  fuel  &  $L$ [m]  &  $R_\mathrm{IBD}$ [$\mathrm{day}^{-1}$]  &  S/B  &  $\sigma_{E,\mathrm{Ph}}/E$ \\
  \hline
  DANSS	~\cite{Danilov:ICHEP2014}		& PS			& 0.9	 & 3000 	& LEU	 & 9.7-12.2 		& $10^4$ 	& 100		& 0.18    \\
  NEOS~\cite{NEOS2015AAP}				& Gd-LS 		& 1	 & 2800   	& LEU 	 & 25  	 		& 1000   	& 22		&  0.05   \\
  Neutrino-4~\cite{Serebrov:2012sq,Serebrov:2013yaa}	& Gd-LS 		& 1.4	 & 90 		& HEU	 & 6-12			& 1800		& $\sim\,$1 	& - 	  \\
  Stereo~\cite{PEQUIGNOT2015126}			& Gd-LS			& 1.8	 & 57		& HEU	 & 8.8-11.2		& 410		& 1.5		& 0.05	  \\
  SoLid	~\cite{Solid2015} 			& PS 			& 2.9	 & 60-80	& HEU	 & 6-8			& 1200 		& 3		& 0.14	  \\
  Prospect I~\cite{Ashenfelter:2015uxt}			& ${}^{6}$Li-LS 	& 1.5 & 85		& HEU	 & 7-12 	& 660	& 3 	& 0.045	  \\
  \hline\hline
  \end{tabular}
\end{table*}
\\
In contrast to this, the Double Chooz experiment observes an antineutrino flux where the number of fissions is shared among ${}^{235}\mathrm{U}$, ${}^{239}\mathrm{Pu}$, ${}^{238}\mathrm{U}$, and ${}^{241}\mathrm{Pu}$ with average fission fractions of 0.49, 0.35, 0.09, and 0.07~\cite{PhysRevD.86.052008}. Variations in the fission fractions $\alpha$ are caused by reactor fuel burn-up. This opens the possibility to search for variations of spectral features linked to changes in the composition of fissioning actinides. Since reactors are partly refueled at different times, they have a complicated burn-up history.
Experiments with multiple reactor cores might therefore be less suitable for this particular measurement, as their detectors measure the integrated flux of all reactors nearby. But the simple experimental configuration of Double Chooz with only two reactor cores allows reactor spectrum measurements at different average fission fractions. Moreover, the best case of a single reactor measurement is given during regular refueling phases with only one detector running.
\\\\
A LEU fueled core will emit an antineutrino spectrum
\begin{equation} \label{eq2} 
S_\mathrm{LEU}(E_{\nu}) = \sum_{k} \alpha_k S_{k}(E_{\nu})\,,
\end{equation}
where $k={}^{235}$U, ${}^{239}$Pu, ${}^{238}$U and ${}^{241}$Pu. The spectra $S_{k}(E_{\nu})$ in Eq.~(\ref{eq2}) are normalized to the number of antineutrinos created per fission by the $k$-th isotope. Then, the fission fraction $\alpha_k$ reflects the portion of fissions provided by the actinide $k$.
\\
As discussed above, the shoulder in the antineutrino spectra might be related to one specific actinide. The antineutrino spectra measured at reactors fueled with LEU or HEU provide the required information by means of their spectral shape. These spectra are e.g.\ measured by the Double Chooz near detector and the Stereo experiment. We will show the potential to test the following three hypotheses: the shoulder is created (1) with similar strength by all actinides, (2) solely by ${}^{235}$U or (3) by any actinide except ${}^{235}$U.
\\\\
Based on the reference antineutrino spectra from Huber-Haag~\cite{PhysRevC.84.024617,haagDC}, expected datasets for the Double Chooz near detector and the Stereo experiment can be computed. The $\bar{\nu}_e$ spectrum observed by Double Chooz is computed using Eq.~(\ref{eq2}) assuming the fission fractions to be as quoted above. For Stereo the reactor neutrino spectrum is taken to correspond to a $\bar{\nu}_e$ spectrum emitted by ${}^{235}$U solely.
In order to introduce the shoulder artificially, a Gaussian shaped excess is added to the spectra. The width and integral of the Gaussian is obtained from the Double Chooz spectrum published in Ref.~\cite{DCIII}.
In our computation, the same percentage of detected IBD events in weight is given to the additional Gaussian of our Double Chooz prediction. The strength of the shoulder in the Stereo prediction is adjusted with respect to the above-mentioned three hypotheses.
\\
In Fig.~\ref{fig1} the event ratio of the Double Chooz near detector and the Stereo prediction is plotted for two calendar years of data taking assuming a duty cycle of about 80\,\% (50\,\%) for Double Chooz (Stereo). The blue solid line is the case where all four isotopes contribute to the event excess with the same strength and is similar to the ratio in absence of any spectral distortion. It is obtained by fits of 5-th order polynomial to the converted Huber-Haag spectra, for the Huber spectra the coefficients can be found in Ref.~\cite{PhysRevC.84.024617}. The slope of the ratio rises with energy, as ${}^{239}$Pu, which emits a softer $\bar{\nu}_e$ spectrum, is only present in LEU fuel in a significant amount. If the shoulder in the spectra are produced only by a subset of actinides, a bump-shaped excess (orange dotted line) or reduction (red dashed line) shows up in the ratio.
The computation of the statistical significance of the event excess is performed in the dedicated energy region from 5 to $\SI{7}{MeV}$. In this way the discrimination power is obtained of distinguishing the case of ${}^{235}$U solely (or not at all) causing the bump from the case of similar contribution of all actinides.
The statistical significance with the assumed two years of run time is 5.9$\,\sigma$ for the scenario of no bump in the ${}^{235}$U spectrum, which would imply the excess is only seen in Double Chooz but not in Stereo. Slightly less significant (4.6$\,\sigma$) would be the case with an event excess in ${}^{235}$U only. The systematics of the conversion spectra and their correlations both in bins and isotopes are taken into account for both scenarios.
\\
The direct comparison of the two datasets leads to a partial cancellation of the model uncertainties as well as normalization uncertainties.
Still, for a more realistic estimate of the significance the influence of the detector response and remaining systematic uncertainties in the spectrum measurements have to be taken into account.
Inclusion of the detector response slightly weakens the significance of the effect as it washes out the bump structure. With about 8\,\% energy resolution in Double Chooz at $\SI{1}{MeV}$ energy and assuming $\sim\,$12\,\% for Stereo at $\SI{3}{MeV}$ we still obtain a significance of 4.9$\,\sigma$ (3.7$\,\sigma$) for the case of no excess in ${}^{235}$U (excess in ${}^{235}$U only). Here we assume the energy resolution scales just with the statistics of the collected photo electrons. For the case of Stereo the actual purely photostatistical response is indeed better.
In addition systematic effects play a role in the experiment due to the smallness of the detector. These effects dominate the response at low energies and are on the other hand suppressed at higher energies~\cite{Lhuillier:Nu2014}. The assumption made for the energy resolution of Stereo is a simplified yet conservative consideration, as it was chosen to not surpass the expected response of the experiment.
\\\\
Other dominant components of the systematic uncertainty will be the precision of the energy scale and the relative normalization of the data-to-data ratio. The latter will take into account differences in the absolute antineutrino rate, e.g.\ from flux uncertainty or detection efficiency, and can be derived from the normalization of the spectra below $\SI{4}{MeV}$ visible energy. The influence of background is negligible for the case of Double Chooz, but might be significant for Stereo.
Nevertheless, current experiments have shown that it is possible to control background at shallow depths~\cite{PhysRevD.93.112006}. Backgrounds from random coincidences can be measured to high precision using the ``off-time window'' method. The correlated background component is determined during reactor-off phases, e.g.\ in Ref.~\cite{PhysRevD.93.112006} is was shown that their correlated backgrounds are mainly caused by cosmogenic events. Future detectors will very likely outperform former short-baseline experiments, the signal-to-background ratio of Stereo is foreseen to be better than for Ref.~\cite{PhysRevD.93.112006}.
\\
Among all sources of systematic uncertainty, in particular the energy scale has a strong impact on the result as shown in the following.
\begin{figure}
	\centering
	\includegraphics[width=0.5\textwidth]{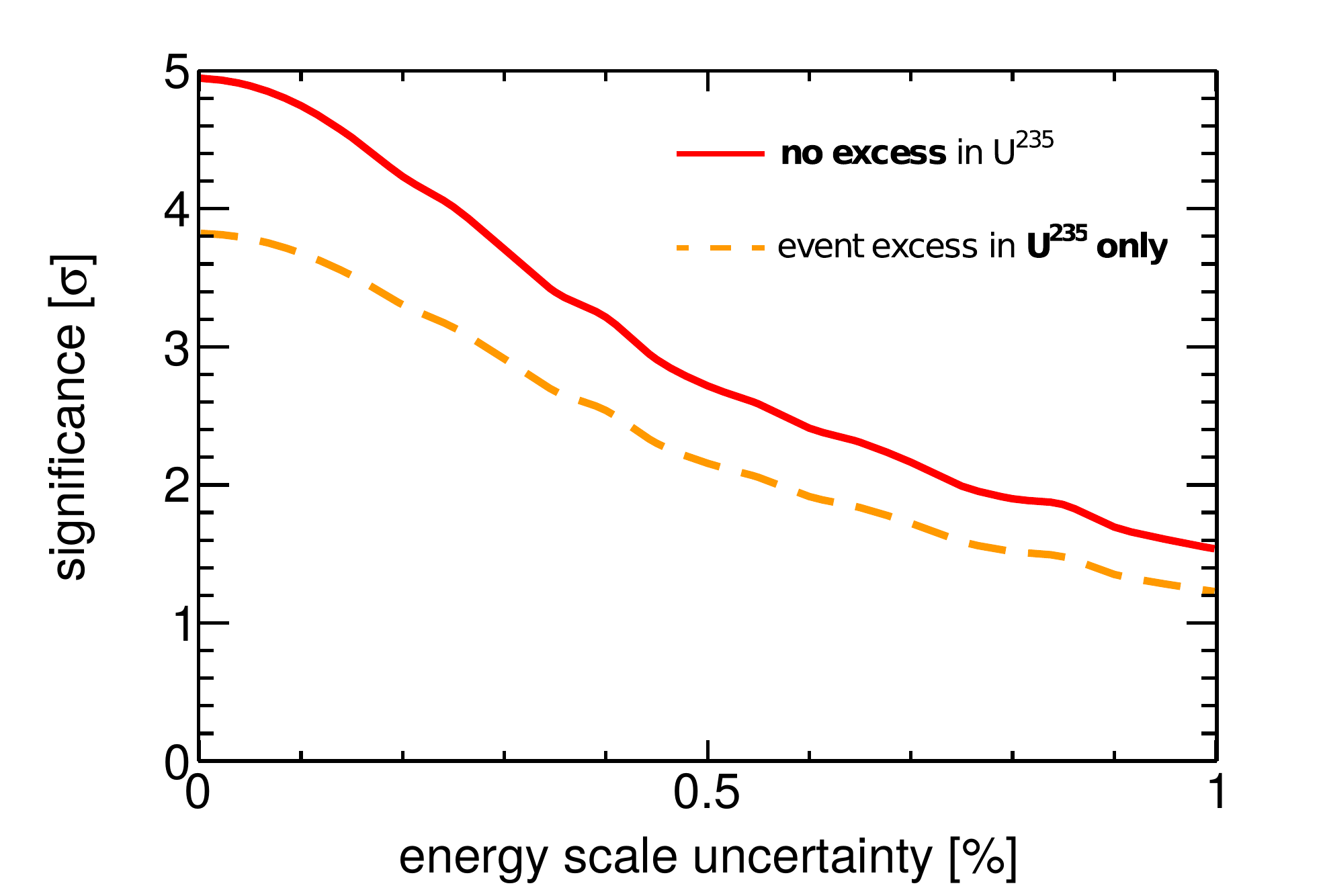}
	\caption{Significance as a function of the energy scale uncertainty.}
	\label{fig2}
\end{figure} 
Therefore, sophisticated and accurate detector calibration also at higher energies above $\SI{5}{MeV}$ is crucial. In Figure~\ref{fig2} the dependence of the significance on the absolute knowledge of the energy scale is shown. The energy scale of each experiment is varied in an additional linear term. Furthermore we assumed the energy scale uncertainty to be the same in both detectors, however fully uncorrelated.
For different energy scale uncertainties the significance will be strongly limited by the detector with the larger error. The precision on the energy scale in the recent $\theta_{13}$-experiments Double Chooz, RENO and Daya Bay is already in the sub per cent regime~\cite{DCIII,PhysRevLett.115.111802,RENO:2015ksa} and might further improve in the future. Whether a similar precision can be reached in a smaller Stereo type detector needs to be demonstrated. In the near future, other short baseline projects will follow, which aim to reach the required energy scale systematics and low background standards~\cite{Ashenfelter:2015uxt}. If such a precision will not be achieved Stereo one might consider to constrain the energy scale systematics using the 2-5$\,\si{MeV}$ range when comparing the experiments to improve the sensitivity.
\\
The remaining uncertainty will come from the statistics in the energy region used for the relative normalization and background systematics. Expecting these influences on the data-to-data ratio to be about 0.5\,\%, the effect on the sensitivity is small, leading to a degradation of the significance of less than 5\,\% for an energy scale uncertainty of $\geq\,$0.3\,\% and 10\,\% at largest if the energy scale uncertainty is lower than 0.3\,\%.
\\\\
Our study shows in summary that the comparison of the measured reactor antineutrino spectrum in the Double Chooz and Stereo experiments is a powerful tool to study the origin of the observed shape distortion compared to flux predictions. After two years of data taking sufficient statistics is collected allowing to distinguish different scenarios as no event excess in the 5-7$\,\si{MeV}$ region in ${}^{235}$U or a distortion exclusively in the ${}^{235}$U spectrum at the 5$\,\sigma$ level, neglecting detection systematics. The sensitivity of such an analysis is driven by the precision of the energy scale, systematic uncertainty contributions of background and flux are expected to be small. Recent liquid scintillator experiments have demonstrated that an energy scale uncertainty well below 1\,\% can be reached, which is the crucial requirement to obtain a significant result.

\section*{Acknowledgements}
The authors would like to thank D.~Lhuillier from CEA Saclay for interesting discussions and valuable comments.
\bibliography{mybibfile}

\end{document}